\documentclass[aps,twocolumn,showpacs,prb]{revtex4-1}
\usepackage{amsmath}
\usepackage{amssymb}
\usepackage{graphicx}
\usepackage{dcolumn}
\usepackage{bm}

\usepackage{bm}
\usepackage{graphicx}
\usepackage{ulem}

\def\be{\begin{equation}}
\def\ee{\end{equation}}
\def\ber{\begin{eqnarray}}
\def\eer{\end{eqnarray}}
\def\bwt{\begin{widetext}}
\def\ewt{\end{widetext}}

\begin{document}

\draft
\title {Effect of nonhomogenous dielectric background on the plasmon modes in\\
graphene double-layer structures at finite temperatures}
\author{S. M. Badalyan}
\email{Samvel.Badalyan@ua.ac.be}
\affiliation{Department of Physics, University of Antwerp, Groenenborgerlaan 171, B-2020 Antwerpen, Belgium}
\author{F. M. Peeters}
\affiliation{Department of Physics, University of Antwerp, Groenenborgerlaan 171, B-2020 Antwerpen, Belgium}

\date{\today}

\begin{abstract}
We have calculated the plasmon modes in graphene double layer structures at finite temperatures, taking into account the inhomogeneity of the dielectric background of the system. The effective dielectric function is obtained from the solution of the Poisson equation of three-layer dielectric medium with the graphene sheets located at the interfaces, separating the different materials. Due to the momentum dispersion of the effective dielectric function, the intra- and inter-layer bare Coulomb interactions in the graphene double layer system acquires an additional momentum dependence--an effect that is of the order of the inter-layer interaction itself. We show that the energies of the in-phase and out-of-phase plasmon modes are determined largely by different values of the spatially dependent effective dielectric function. The effect of the dielectric inhomogeneity increases with temperature and even at high temperatures the energy shift induced by the dielectric inhomogeneity and temperature itself remains larger than the broadening of the plasmon energy dispersions due to the Landau damping. The obtained new features of the plasmon dispersions can be observed in frictional drag measurements and in inelastic light scattering and electron energy-loss spectroscopies.
\end{abstract}


\maketitle

\section{Introduction}
Graphene is a monolayer of carbon atoms \cite{graphene,wallace,geimnovo2007} with great potential for a new generation electronics \cite{geim2009,geimmac}. Charge carriers in graphene are Dirac-like massless, chiral fermions that provide a unique two dimensional system with new many-body phenomena \cite{kotov} that can critically influence the electronic properties \cite{Yan,Yuan,abedinpour,plasmaron,hwang1,Jang,Barlas,Cheianov} in graphene. 
An excellent tool for studying many-body interaction in graphene structures are graphene double-layer systems (GDLS), recently realized in several experiments \cite{falko,tutuc,ponamarenko}.

Bare electron-electron interaction in graphene is described by the dimensionless fine-structure constant \cite{Nair}
\begin{equation}
\label{alpha}
\alpha_{g}=\frac{e^{2}}{\hbar  v_{F}\bar{\epsilon}} \approx \frac{2.2}{\bar{\epsilon}}
\end{equation}
with its value depending on the  dielectric properties of the graphene surrounding environment via the effective dielectric constant $\bar{\epsilon} $ ($v_{F}$ the electron velocity in graphene). In GDLS the two spatially separated graphene sheets are immersed in a nonhomogeneous three layered medium with background dielectric constants $\epsilon_{1}$, $\epsilon_{2}$ and $\epsilon_{3}$ of the contacting media, as shown in Fig.~1. 
In general, these dielectric permittivities differ substantially from each others in experiment. In an individual graphene sheet on top of a substrate with relative dielectric permittivity $\epsilon_{d}$, an electron charge $e$ behaves effectively as a charge with a value of $2e/(1+\epsilon_{d})$. This well known result \cite{landau} has been applied in recent treatment of Coulomb drag in GDLS \cite{castroneto}, assuming that the effective permittivity in each graphene layer of the GDLS is given by the arithmetic average of its surrounding media, i.e. $\bar{\epsilon}_{12}=(\epsilon_{1}+\epsilon_{2})/2$ and $\bar{\epsilon}_{23}=(\epsilon_{2}+\epsilon_{3})/2$. In what follows, we argue that such an approach in general is not applicable for the plasmon problem in GDLS. It neglects the momentum dispersion of the effective dielectric permittivity, $\bar{\epsilon}=\bar{\epsilon}(q)$, an effect of the order of the inter-layer interaction. Its direct application would result in the disappearance of the linear dispersion of out-of-phase plasmon modes  because the bare intra- and inter-layer Coulomb interactions remain not equal in the long wavelength limit. Furthermore, when an average spatially independent dielectric permittivity is used \cite{sarma,gumbs,parhizgar,khanh} to describe GDLS,  we find that it also provides an inadequate description of the plasmon energy dispersions in GDLS. In the present paper we develop a consistent description of the plasmon modes in GDLS by making use of the exact solution of the Poisson equation for the electrostatic problem in three layer dielectric medium and by taking into account the momentum dispersion of the effective background dielectric function $\bar{\epsilon}(q)$ due to the finite thickness of the inter-layer barrier. 
\begin{figure}[h]
\includegraphics[width=4cm]{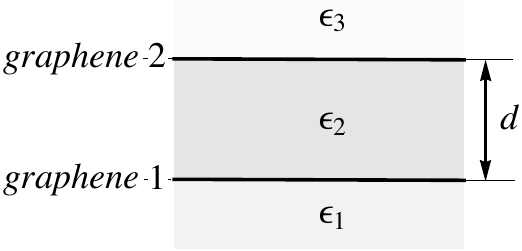}
\caption{A graphene double-layer system immersed in a three layered dielectric medium. The solid lines with spacing $d$ represent the graphene sheets $1$ and $2$ at the interfaces, separating different materials with the background dielectric permittivities $\epsilon_{1}$, $\epsilon_{2}$, and $\epsilon_{3}$. }
\label{fig1}
\end{figure}
We show that even in the long wavelength limit, when the bare intra- and inter-layer Coulomb interactions remain equal, the dispersions of the in-phase and out-of-phase plasmon modes are determined by the largely different values of the spatially dependent effective dielectric permittivity. At finite momenta also the momentum dependence of the effective dielectric function $\bar{\epsilon}(q)$ results in significant changes of the dispersion relation of both plasmon modes which can be probed in experiment. 

\section{Theoretical concept}
We obtain the collective modes of plasmon excitations \cite{GV} from the poles of the exact Coulomb Green function, $\hat{V}(q,\omega)$. In double layer structures the bare Coulomb interaction, $\hat{v}(q)$, is a tensor with respect to the layer indices and in general represents three different interactions, the intra-layer, $v_{11}(q)$ and $v_{22}(q)$, and the inter-layer, $v_{12}(q)=v_{21}(q)$. The kernel $\hat{v}(q)$ determines a standard matrix Dyson equation for the exact Coulomb Green function
\begin{equation}
\hat{V}(q,\omega)=\hat{v}(q)+\hat{v}(q)\cdot\hat{\Pi}(q,\omega)\cdot\hat{V}(q,\omega)
\label{dyson}
\end{equation}
where $\hat{\Pi}(q,\omega)$ is the irreducible polarization function of the double layer electron system. It is seen from (\ref{dyson}) that the poles of the exact Coulomb Green functions are given by the zeros of the scalar screening function $\varepsilon(q,\omega)=\text{det} |1-\hat{v}(q)\cdot\hat{\Pi}(q,\omega)|$.  In GDLS even for the inter-layer spacing of 2-3~nm, the Dirac carriers in the graphene sheets are coupled via inter-layer Coulomb interaction and the tunneling between the layers is insignificant. Hence we can neglect the non-diagonal elements of the polarizability $\hat{\Pi}(q,\omega)$. It is also sufficient to restrict ourselves to consider the polarizability within the random phase approximation where $\hat{\Pi}_{11}(q,\omega)=\Pi^{0}_{1}(q,\omega)$ and $\hat{\Pi}_{22}(q,\omega)=\Pi^{0}_{2}(q,\omega)$ and the noninteracting polarization functions $\Pi^{0}_{1,2}(q,\omega)$ are given by the bubble diagrams in the respective graphene sheets. This approach is well justified for weakly interacting GDLS and its screening function can be represented as 
\begin{equation}
\varepsilon(q,\omega)=\varepsilon_{1}(q,\omega)\varepsilon_{2}(q,\omega)-v_{12}(q)^{2}\Pi^{0}_{1}(q,\omega)\Pi^{0}_{2}(q,\omega)
\label{screening}
\end{equation}
where $\varepsilon_{1,2}(q,\omega)=1-v_{11,22}(q)\Pi^{0}_{1,2}(q,\omega)$ are the screening functions in each graphene layer. Despite the external similarities, the screening function of GDLS differs essentially from that of the usual two-dimensional electron gas in semiconductor nanostructures. %with parabolic dispersion law of carriers. 
In addition to the new properties of $\Pi^{0}(q,\omega)$ arising from the unique Dirac-like energy bandstructure and the chiral nature of the massless carriers, we show here that the heterogeneity of the dielectric background in GDLS plays an important role in determining the many-body Coulomb interaction effects. 

\begin{figure*}[t]
\includegraphics[width=5.75cm]{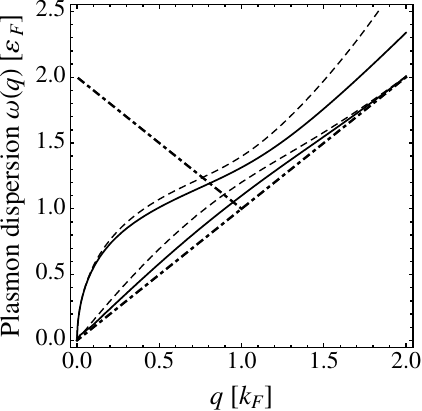}\hfill\includegraphics[width=5.75cm]{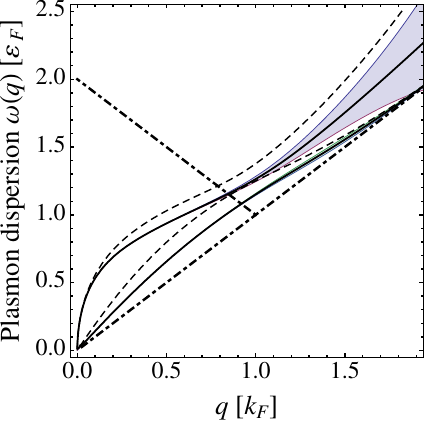}\hfill\includegraphics[width=5.75cm]{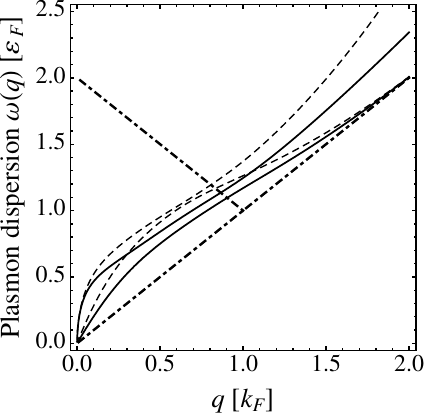}
\caption{The effect of  nonhomogenous dielectric background on the plasmon dispersions in GDLS at low temperatures $T=0.1T_{F}$. The upper (lower) curves represent the optical (acoustical) plasmon branches with the square-root (linear) dispersions. The solid curves are calculated for the plasmon modes in GDLS with nonhomogenous dielectric background, consisting of three layers of SiO$_{2}$ ($\epsilon_{{SiO_{2}}}=3.8$), Al$_{2}$O$_{3}$ ($\epsilon_{Al_{2}O_{3}}=6$), and air. The dashed curves correspond to the plasmons in GDLS with homogenous dielectric background with an average permittivity $\epsilon=2.4$. In the left, mid, and right figures we plot, respectively, the plasmon energy dispersions for three values of the inter-layer spacing $d=5$ nm, $10$ nm, and $30$ nm. The doping level corresponds to the carrier densities $n_{1}=n_{2}=10^{12}$ cm$^{-2}$.  The thick dot-dashed lines show the boundaries of inter- and intra-chirality particle-hole continua where plasmons are Landau damped. The shaded areas in the mid plot represent the broadening of the energy dispersions of respective plasmon modes.}
\label{fg2}
\end{figure*}

Direct calculations of the electrostatic problem in the dielectric environment consisting of three contacting media with different dielectric constants $\epsilon_{1}$, $\epsilon_{2}$, and $\epsilon_{3}$ give the following formula for the Coulomb potential
\begin{widetext}
\begin{eqnarray}
\varphi(z,z_{0})&=&\frac{2\pi e^{2}}{q\epsilon_{2}}
\left(e^{-q | z-z_{0} |}\right. \nonumber \\ 
&+&
\left.
\frac{\left(\epsilon_{2}-\epsilon_{3} \right) \left(\epsilon_{1} \sinh q z_{0} + \epsilon_{2}\cosh q z_{0} \right) e^{q (z-d)}
-\left(\epsilon_{1}-\epsilon_{2}\right) \left(\epsilon_{2} \cosh q (d-z_{0}) + \epsilon_{3} \sinh q (d-z_{0}) \right)e^{-q z}}
{\epsilon_{2} \left(\epsilon_{1}+\epsilon_{3}\right) \cosh d q + \left(\epsilon_{1} \epsilon_{3} + \epsilon_{2}^2 \right)
\sinh qd } \right)
\end{eqnarray}
\end{widetext}
where $z_{0}$ and $z$ are, respectively, the positions of the source and test charges. The bare intra-layer interactions in graphene layers $1$ and $2$, located at the two interfaces with $z=0$ and $z=d$, are given by $v_{11}(q)=e \varphi(0,0)$ and $v_{22}=e \varphi(d,d)$ while the bare inter-layer interaction $v_{12}(q)=e \varphi(0,d)$ \cite{fil,Profumo0}. Hence, we arrive at the following three effective dielectric functions in the GDLS
\begin{eqnarray}
\frac{1}{\bar{\epsilon}_{11}(q )} =
\frac{2\left(\epsilon_{2} \cosh q d + \epsilon_{3} \sinh q d \right)}
{\epsilon_{2} \left(\epsilon_{1}+\epsilon_{3}\right) \cosh d q + \left(\epsilon_{1} \epsilon_{3} + \epsilon_{2}^2 \right)
\sinh qd } \label{effdiel1} \\
\frac{1}{\bar{\epsilon}_{22}(q )} =
\frac{2\left(\epsilon_{2} \cosh q d + \epsilon_{1} \sinh q d \right)}
{\epsilon_{2} \left(\epsilon_{1}+\epsilon_{3}\right) \cosh d q + \left(\epsilon_{1} \epsilon_{3} + \epsilon_{2}^2 \right)
\sinh qd } \label{effdiel2} \\
\frac{1}{\bar{\epsilon}_{12}(q )} =
\frac{2 \epsilon_{2}} %e^{q d}
{\epsilon_{2} \left(\epsilon_{1}+\epsilon_{3}\right) \cosh d q + \left(\epsilon_{1} \epsilon_{3} + \epsilon_{2}^2 \right)
\sinh qd }
\label{effdiel3}
\end{eqnarray}
that determine the strength of the intra- and inter-layer bare Coulomb interactions in momentum space
\begin{eqnarray}\label{renormint}
v_{ij}(q,d)=\frac{2\pi e^{2}}{q \bar{\epsilon}_{ij}(q d)}
\end{eqnarray}
Here $i, j = 1,2$ are the graphene layer indices. It is seen that in the long wavelength limit all three interactions are determined by the same effective dielectric constant, given by the arithmetic average of the top and bottom surrounding media in GDLS, $\bar{\epsilon}_{13}=(\epsilon_{1}+\epsilon_{3})/2$, and does not depend on the dielectric constant, $\epsilon_{2}$, of the middle medium. 

Further we exploit the above formulae for bare interaction $v_{ij}(q)$ to find the plasmon spectrum in GDLS.
Assuming for simplicity that the density is balanced in the GDLS, we can rewrite the screening function (\ref{screening}) as
\begin{eqnarray}
\varepsilon(q,\omega)=a\left(\Pi^{0}(q,\omega)-\Pi^{+}(q,d)\right)\left(\Pi^{0}(q,\omega)-\Pi^{-}(q)\right)
\label{screening2}
\end{eqnarray}
where $a(q,d)=v_{11}(q,d)v_{22}(q,d)-v_{12}^{2}(q,d)$ and the auxiliary functions
\begin{equation}
\Pi^{\pm}(q,d)=\frac{v_{11}+v_{22} \pm \sqrt{( v_{11}-v_{22})^{2}+4v_{12}^{2}}}{2a(q,d)}~.
\end{equation}
Here for brevity we omit the arguments of the bare interactions. In the limit of vanishing $q$ we have
\begin{equation}\label{aux}
\Pi^{+}(q,d)=\frac{\bar{\epsilon}_{13}}{4\pi e^{2}}q~,~\Pi^{-}(q,d)=\frac{\epsilon_{2}}{2\pi e^{2} d}~.
\end{equation}
In unbalanced GDLS instead of the above formulas, one can find
\begin{equation}
\Pi^{+}(q)=\frac{\bar{\epsilon}_{13}}{2\pi e^{2}}\frac{q}{\sqrt{n_{1}}+\sqrt{n_{2}}}~,~\Pi^{-}(d)=\frac{\epsilon_{2}}{2\pi e^{2}}\frac{\sqrt{n_{1}}+\sqrt{n_{2}}}{2 d \sqrt{n_{1} n_{2}}}
\end{equation}
and the dielectric inhomogeneity has a similar effect also in unbalanced systems. Here the partial dimensionless densities are defined as $n_{1,2}=N_{1,2}/N$ with $N_{1,2}$ the carrier densities in each graphene layers and $N=N_{1}+N_{2}$ stands for the total density in GDLS.

Making use of the expression (8) from Ref.~\onlinecite{HDS2007} for the zero temperature exact Lindhard polarization function in graphene, we find from the zeros of the screening function in Eq.~(\ref{screening2}) the optical and acoustical plasmon modes in GDLS with the following square-root and linear energy dispersions
\begin{equation}\label{DR}
\omega_{+}(q)=\sqrt{\frac{g e^{2} v_{F} k_{F}}{\bar{\epsilon}_{13}}q}~,~\omega_{-}(q)=
\frac{1+ q_{TF}d}{\sqrt{1+ 2 q_{TF}d}} v_{F}q~.
%\frac{1+ g \alpha_{2} k_{F}d}{\sqrt{1+ 2 g \alpha_{2} dk_{F}}}v_{F}q~.\\
\end{equation}
Here $g$ accounts for the spin and valley degeneracy in graphene and $q_{TF}=g e^{2}k_{F}^{2}/(\epsilon_{2}E_{F})$ is the Thomas-Fermi screening wave vector in graphene with $k_{F}$ defined by the single layer density. 
These plasmon dispersions, derived for GDLS in Ref.~\onlinecite{Profumo} (see the {\it Note added}), have the same form as for semiconductor two dimensional systems \cite{flensberg}. It is seen, however, that in contrast to the previous treatments, adopting a model of homogenous background dielectric environment for the GDLS \cite{sarma,gumbs,parhizgar,khanh}, the plasmon energies (\ref{DR}) of the optical and acoustical modes in GDLS are determined  by the different background dielectric permittivities, which is a direct consequence of the behavior of the auxiliary functions from (\ref{aux}) in the long wavelength limit.

\section{Results and discussions}

It follows from the above formulas that the energy of the in-phase optical plasmons is given by the arithmetic average $\bar{\epsilon}_{13}$ of the dielectric constants of top and bottom dielectric media in the GDLS and is independent of the middle layer permittivity, $\epsilon_{2}$, while the energy of the out-of-phase acoustical plasmon modes is independent of $\epsilon_{1},~\epsilon_{3}$ and depends only on the dielectric constant $\epsilon_{2}$. 
In recent experimental samples of GDLS \cite{falko,tutuc,ponamarenko} the top graphene sheet is surrounded by air while the inter-layer barrier has a relatively large dielectric permittivity. In samples where the bottom graphene layer in GDLS lies on a boron nitride substrate, used in the experimental setup in Ref.~\onlinecite{ponamarenko}, $\epsilon_{1}=\epsilon_{2}=\epsilon_{\text{BN}}\approx5$ %and $\epsilon_{3}=\epsilon_{\text{air}}$
%\begin{eqnarray}\epsilon_{1}=\epsilon_{BN}=5~,~\\\epsilon_{2}=\epsilon_{BN}=5~,~\\\epsilon_{3}=\epsilon_{air}=1~,~\end{eqnarray}
and we have $\bar{\epsilon}_{13}\approx 3$, which is by a factor $1.7$ smaller than $\epsilon_{2}=5$.
%\begin{eqnarray}\epsilon_{1}=\epsilon_{{SiO_{2}}}=3.8~,~\\\epsilon_{2}=\epsilon_{Al_{2}O_{3}}=6~,~\\\epsilon_{3}=\epsilon_{\text{air}}=1~,~\end{eqnarray}
In samples on a SiO$_{2}$ substrate with moderate dielectric effects, studied in Ref.~\onlinecite{falko,tutuc} $\epsilon_{1}=\epsilon_{\text{SiO}_{2}}\approx 3.8$ and $\epsilon_{2}=\epsilon_{\text{Al}_{2}\text{O}_{3}}\approx 6$ %and $\epsilon_{3}=\epsilon_{\text{air}}$ 
and we have $\bar{\epsilon}_{13}\approx 2.4$, which is smaller than $\epsilon_{2}=6$ by a factor of about $2.5$.

\begin{figure*}[t]
\includegraphics[width=5.75cm]{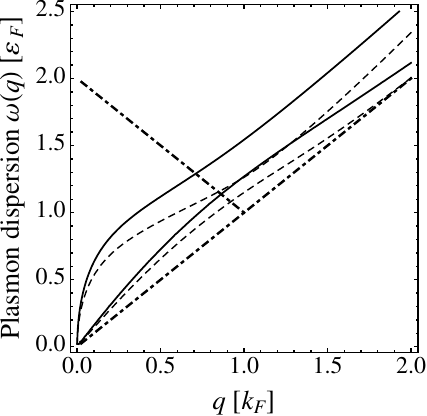}\hfill\includegraphics[width=5.75cm]{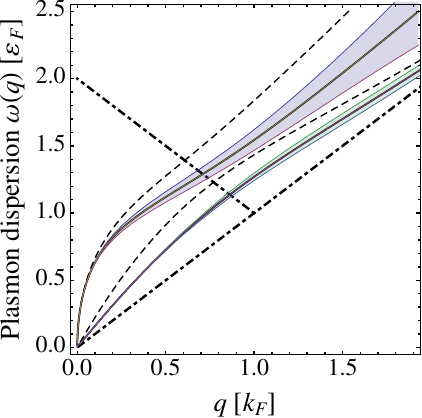}\hfill\includegraphics[width=5.75cm]{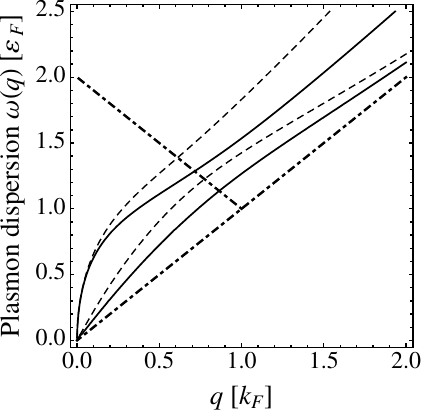}
\caption{(left) Temperature effect on the in-phase and out-of-phase plasmon modes in GDLS with nonhomogenous dielectric background. The dashed and solid lines correspond to the $T=0.1T_{F}$ and $T=T_{F}$.
(mid) and (right) The effect of nonhomogenous dielectric background on the plasmon dispersions in GDLS at high temperature $T=T_{F}$. The plasmon dispersions are shown (mid) in the exactly balanced, $n_{1}=n_{2}=10^{12}$ cm$^{-2}$, and (right) in the completely unbalanced, $n_{2}=0$, GDLS. In all figures $d=10$ nm and other parameters and notations correspond to Fig.~\ref{fg2}. The shaded areas in the mid plot represent the broadening of the energy dispersions of respective plasmon modes.}
\label{fg5}
\end{figure*}

Thus, the inhomogeneity of the background dielectric environment in GDLS can significantly alter the dispersions of plasmon modes in the long wavelength limit by reducing essentially the energy of the out-of-phase plasmon branch (see numerical calculations below). 
In particular, this can lead to strong modifications of the temperature dependence of the plasmon-mediated Coulomb drag in GDLS \cite{katsnelson,hwang,tse,castroneto}. At low temperature the plasmon-mediated drag is mainly determined by the acoustical plasmon modes, which will be easy to excite thermally because they become closer to the boundary of the particle-hole continuum. As a result the upturn temperature of the plasmon enhancement will be shifted to lower temperatures with the higher enhancement peak of the drag resistivity.

The momentum dispersion that the effective dielectric functions $\bar{\epsilon}_{ij}(q)$ exhibit in Eqs.~(\ref{effdiel1}-\ref{effdiel3}) modifies substantially also the double-layer plasmon dispersions at finite values of $q$. In Figs.~\ref{fg2} and \ref{fg5} we calculate the plasmon dispersions at two different temperatures $T=0.1T_{F}$ and $T=T_{F}$ in the range of momenta, $0<q<2k_{F}$, from the zeros of the screening function (\ref{screening}) by making use of the exact semi-analytical formulas from Ref.~\onlinecite{ramez} for the finite temperature polarization function $\Pi_{i}^{0}(q,\omega | T)$ of graphene. 

As seen in Fig.~\ref{fg2} the account for the inhomogeneity of the dielectric background in GDLS suppresses strongly the out-of-phase acoustical plasmon mode. The largest energy difference in comparison with the homogenous case is achieved nearly at the boundary of the inter-chirality particle-hole continuum for $q\sim k_{F}$. The effect of inhomogeneity of the dielectric background on the energy of in-phase plasmon modes starts to be significant at finite values of $q\sim 0.2k_{F}$. It increases with $q$ so that the dispersion curve remains approximately parallel to the boundary of the intra-chirality subband particle-hole continuum at momenta $q\sim 2k_{F}$. In contrast, the energy deviation of the out-of-phase plasmon modes due to the dielectric inhomogeneity shows nonmonotonic behavior and for large values of $q\sim k_{F}$ starts to decrease with $q$. It is seen also that the dispersion curves of the in-phase and out-of-phase plasmon modes exhibit an usual behavior and with an increase of the inter-layer spacing two modes become degenerate outside the particle-hole continua. The shaded areas in the middle figures of Figs.~\ref{fg2} and \ref{fg5} represent the broadening of the respective plasmon dispersions in the complex $\omega$ plane. Instead of the absolute value of $\Im \Pi^{0}(q,\omega)$, the broadening $\Gamma_{\pm}(q)$ is the real measure of the plasmon Landau damping and is given by \cite{flensberg} 
\begin{eqnarray}
\Gamma_{\pm}(q) =\left. \frac{ \Im \Pi^{0}(q,\omega)} {\partial \Re \Pi^{0}(q,\omega) / \partial\omega}\right|_{\omega=\omega_{\pm}(q)}
\end{eqnarray}
Far from the boundaries of the particle-hole continua, the plasmon modes are strongly Landau damped and the dispersion curves do not represent well defined elementary excitations. With an increase of $T$ the particle-hole continua ``degrade'' while the damping of plasmon modes with smaller values of $q$ increases. It is seen, however, that even at high temperatures $T=T_{F}$ the broadening of the plasmon dispersions for $q\lesssim k_{F}$ is smaller than the shift of the energy dispersions, induced by the combined action of the dielectric background inhomogeneity and temperature. This is true both for the optical and acoustical plasmon modes. In this small $q$ and small $\omega$ region $\Im \Pi^{0}(\omega,q)$ is small and because its contribution to the real part of the screening function in Eq.~(\ref{screening}) is quadratic, the plasmon dispersion relations obtained from the zeros of the factors of the screening function in (\ref{screening2}) give the same result against the background of the broadening of plasmon dispersions, linearly varying  with $\Im \Pi^{0}(q,\omega)$.

In Figs.~\ref{fg5} we study the combined effect of finite temperatures and the inhomogeneous dielectric background on the energy dispersions of double-layer plasmon modes. It is seen in Fig.~\ref{fg5}(left) that the plasmon energy of both modes increases significantly with $T$. The comparison of Figs.~\ref{fg2}(mid) and \ref{fg5}(mid) shows that the effect of the dielectric background inhomogeneity becomes stronger with $T$. As seen from Figs.~\ref{fg5}(mid) and \ref{fg5}(right) at finite temperatures $T=T_{F}$ the plasmon modes behave almost in the same way in two extreme regimes of the completely balanced, $n_{1}=n_{2}$, and the completely unbalanced, $n_{2}=0$, GDLS. This is in stark contrast to the behavior that plasmons exhibit at $T=0$ \cite{sarma}. At the same time Fig.~\ref{fg5}(right) shows that the effect of dielectric background inhomogeneity in the unbalanced GDLS is similarly strong as in the balanced GDLS. 

\section{Summary}
We study the effect of the dielectric background inhomogeneity on the plasmon modes in GDLS at finite temperatures. It is found that using a spatially averaged dielectric permittivity to describe dielectric properties of GDLS provides an inadequate description of the plasmon energy dispersions. We obtain the effective dielectric functions of GDLS from the exact solution of the Poisson equation for a three-layer dielectric medium with the graphene sheets located at the interfaces, separating different materials. It is shown that the momentum dispersion of the effective dielectric function results in an additional momentum dependence of the intra- and inter-layer bare Coulomb interactions in GDLS--an effect of the order of inter-layer interaction itself. As a result the obtained dispersions of the in-phase and out-of-phase plasmon modes are largely determined by the different values of the spatially dependent effective dielectric permittivity. 

Here we use the temperature dependent exact Lindhard polarization function, which allows us to find the plasmon dispersions at finite temperatures. Our calculations show that the effect of dielectric background inhomogeneity increases with temperature and even at high temperatures, $T\sim T_{F}$, the energy shift induced by the dielectric inhomogeneity and temperature itself remains larger than the broadening of the plasmon energy dispersions due to the Landau damping at finite temperatures. This combined effect of finite temperatures and the dielectric background inhomogeneity has been discussed thoroughly in our paper by comparing carefully the dispersion relations, calculated for graphene double layers embedded in a three-layer nonhomogeneous dielectric medium with that obtained for the homogeneous one. We find a strong effect of temperatures on the acoustical plasmon mode in a completely unbalanced system--the acoustical mode becomes separated from the top of the electron-hole continuum and this favors possibility to observe it. Our numerical calculations have been carried out for realistic experimental samples on a Si substrate with three different dielectric constants for the background nonhomogeneous medium for interlayer spacing $d=5, 10$, and $30$ nm. The effect of dielectric inhomogeneity is found to be the largest for samples with $d=10$ nm at finite momenta $q\lesssim k_F$.

The predicted new features of the double-layer plasmon dispersions can be observed in high temperature measurements of frictional drag in GDLS and in such plasmon experiments as plasmon-enhanced photoluminescence \cite{Hwang2}, electron energy-loss spectroscopies \cite{Lu,Liu}, inelastic light scattering \cite{Liu,Eberlein} measurements.

{\it Note added }$-$ During the preparation of our manuscript when its main results have been already summarized in the abstract submitted to the March Meeting \cite{mm}, we became aware of two other preprints  (Refs.~\onlinecite{Profumo,Stauber2011}) on the double-layer plasmon modes in graphene structures. In the present paper we give particular emphasis to the effect of inhomogeneity of the dielectric background on the plasmon dispersions in realistic GDLS and use a different formalism from that used in Refs.~\onlinecite{Profumo,Stauber2011}, which allows us to treat additionally the effect of finite temperatures on the energy dispersions of the plasmon modes. The main focus of Ref.~\onlinecite{Profumo} has been the subtle point related to the use of the {\it exact} Lindhard polarization function in obtaining the plasmon velocity in the long wavelength limit. The authors derived first the analytical formula for the acoustical plasmon velocity thereby correcting the previously reported result from Ref.~\onlinecite{sarma}. The effect of dielectric inhomogeneity, which is due to the renormalization of the bare Coulomb interactions, in Ref.~\onlinecite{Profumo} has been discussed mainly in connection with the effect of the locked acoustical mode on top of the particle-hole continuum in topological insulators. We find, however, that the numerical calculations presented in Figs. 2-4 of Ref.~\onlinecite{Profumo} for the interlayer spacing $d=0.335$ nm are not adequate. For such a small value of the barrier thickness the tunneling between graphene layers cannot be neglected and the Eq.~(8) in Ref.~\onlinecite{Profumo} is inapplicable. Although the authors of Ref.~\onlinecite{Stauber2011} have reported the main formulas for the three layer dielectric medium, their main focus has been the study of the near field amplification in GDLS with a homogenous dielectric background.

\section{Acknowledgements}
We thank G. Vignale for useful discussions and acknowledge support from the Flemisch Science Foundation (FWO-Fl) and the Belgian Science Policy (BELSPO).

\end{document}